\documentclass{aa}

\usepackage[varg]{txfonts}
\usepackage{textcomp}
\usepackage{graphicx}
\usepackage{natbib}
\bibpunct{(}{)}{;}{a}{}{,}

\begin{document}
\title{Experimental constraints on the ordinary chondrite shock darkening caused by asteroid collisions}

\author{T.~Kohout\inst{\ref{inst1},\ref{inst2}}\and E.~V.~Petrova\inst{\ref{inst3}}\and G.~A.~Yakovlev\inst{\ref{inst3}}\and V.~I.~Grokhovsky\inst{\ref{inst3}}\and A.~Penttil\"{a}\inst{\ref{inst4}}\and A.~Maturilli\inst{\ref{inst5}}\and J.-G.~Moreau\inst{\ref{inst1}}\and S.~V.~Berzin\inst{\ref{inst6}}\and J.~Wasiljeff\inst{\ref{inst1}}\and I.~A.~Danilenko\inst{\ref{inst3},\ref{inst6}}\and D.~A.~Zamyatin\inst{\ref{inst3},\ref{inst6}}\and R.~F.~Muftakhetdinova\inst{\ref{inst3}}\and M.~Heikkil\"{a}\inst{\ref{inst7}}}

\institute{Department of Geosciences and Geography, University of Helsinki, Finland \email{tomas.kohout@helsinki.fi}\label{inst1}
\and Institute of Geology, The Czech Academy of Sciences\label{inst2}
\and Institute of Physics and Technology, Ural Federal University, Ekaterinburg, Russian Federation\label{inst3}
\and Department of Physics, University of Helsinki, Finland\label{inst4}
\and Institute of Planetary Research, DLR, Berlin, Germany\label{inst5}
\and Zavaritsky Institute of Geology and Geochemistry, Ural Branch of the Russian Academy of Sciences, Ekaterinburg, Russian Federation\label{inst6}
\and Department of Chemistry, University of Helsinki, Finland\label{inst7}}

\abstract
{
Shock-induced changes in ordinary chondrite meteorites related to impacts or planetary collisions are known to be capable of altering their optical properties. Thus, one can hypothesize that a significant portion of the ordinary chondrite material may be hidden within the observed dark C/X asteroid population.
}
{
The exact pressure-temperature conditions of the shock-induced darkening are not well constrained. Thus, we experimentally investigate the gradual changes in the chondrite material optical properties as a function of the shock pressure.
}
{
A spherical shock experiment with Chelyabinsk LL5 was performed in order to study the changes in its optical properties. The spherical shock experiment geometry allows for a gradual increase of shock pressure  from $\sim$15 GPa at a rim toward hundreds of gigapascals in the center.
}
{
Four distinct zones were observed with an increasing shock load.  The optical changes are minimal up to $\sim$50 GPa. In the region of $\sim$50--60 GPa, shock darkening occurs due to the troilite melt infusion into silicates. This process abruptly ceases at pressures of $\sim$60 GPa due to an onset of silicate melting. At pressures higher than $\sim$150 GPa, recrystallization occurs and is associated with a second-stage shock darkening due to fine troilite-metal eutectic grains. The shock darkening affects the ultraviolet, visible, and near-infrared (UV, VIS, and NIR) region while changes to the MIR spectrum are minimal.
}
{
Shock darkening is caused by two distinct mechanisms with characteristic pressure regions, which are separated by an interval where the darkening ceases. This implies a reduced amount of shock-darkened material produced during the asteroid collisions.
}

\keywords{Meteorites, meteors, meteoroids -- Minor planets, asteroids: general -- Shock waves} 
\titlerunning{Ordinary chondrite shock darkening experiment}

\maketitle

\section{Introduction}

Shock-induced changes in planetary materials related to impacts or planetary collisions are known to be capable of altering their optical properties. One such example is observed in ordinary chondrite meteorites. The highly shocked silicate-rich ordinary chondrite material is optically darkened and its typical S-complex-like asteroid spectrum \citep{RN83} is altered toward a darker, featureless spectrum resembling the C/X complex asteroids \citep{RN23, RN24, RN4, RN139}. Thus, one can hypothesize that a significant portion of the ordinary chondrite material may be hidden within the observed C/X asteroid population. The exact pressure-temperature conditions of the shock-induced darkening are, however, not well constrained and due to this gap in knowledge, it is not possible to correctly assess the significance of the shock darkening within the asteroid population. In order to address this shortcoming, we experimentally investigate the gradual changes in the chondrite material optical properties together with the associated mineral and textural features as a function of the shock pressure. For this purpose, we use a Chelyabinsk meteorite (LL5 chondrite), which is subjected to a spherical shock experiment. The spherical shock experiment geometry allows for a gradual increase in the shock pressure within a single spherically shaped sample from $\sim$15 GPa at its rim toward hundreds of gigapascals in the center.

In this study we link the experimentally produced shock features in the Chelyabinsk meteorite to the modeled shock conditions and to the naturally occurring shock features observed in the Chelyabinsk and other chondritic meteorites. For this purpose, we briefly describe the Chelyabinsk meteorites and review the features causing shock darkening.

The Chelyabinsk meteorite fall occurred on February 15, 2013 in Russia, close to the city of Chelyabinsk. The meteorite was classified as LL5 ordinary chondrite \citep{RN81}. Three distinct lithologies were described to be present among collected meteorites \citep{RN4}. The ''light-colored'' lithology is moderately shocked (C-S3) with undulatory extinction in olivine, which is the main diagnostic feature. Mosaicism is observed locally, which is suggestive of a shock stage C-S4. The ''dark-colored'' lithology is characterized by its dark appearance caused by an extensive troilite melt production and its migration into the extra- and intra-granular fractures within the solid silicate fraction. The microscopic classification of this lithology is difficult due to its opaque nature. The extensive troilite melting, however, requires pressures within the shock stage C-S5 \citep{RN458}. The ''impact melt'' lithology appears either as melt pockets or shock veins, dominating fragments of light-colored lithology, or as a more massive body containing floating light-colored and dark-colored clasts \citep{RN962}. Its shock stage corresponds to C-S6 (veins, pockets-dominated material) or C-S7 (massive melt body). At a first glance, the impact melt lithology is similar in its visual appearance to the dark-colored one. The microscopic observations, however, reveal the complete whole-rock melting. Iron-metal and troilite form eutectic blebs within the silicate melt \citep{RN437}. The melt contains varying amounts of finely brecciated residual solid clasts. In some of the samples, new  long-prismatic olivine crystals are observed within the quenched melt \citep{RN914}.

X-ray diffraction (XRD) and Raman spectroscopy analyses indicate that the overall mineralogy of the three Chelyabinsk lithologies is similar \citep{RN913, RN61, RN404}. Newly crystalized long-prismatic olivine crystals, which are more magnesium-rich as described in \cite{RN914}, are an exception to this.

Optical changes in the chondritic materials are associated with a partial or complete melting caused by the impact-related post-shock heating \citep{RN72, RN74, RN77, RN75, RN76}. Shock modeling results suggest that at $\sim$40-50 GPa, troilite starts to extensively melt due to its lower melting point with only a small amount of silicate or metal melt localized in the thin shock veins \citep{RN135, RN458}. The molten troilite typically forms a web of fine veins dispersed in the cracks within solid silicate grains, which is similar to the Chelyabinsk dark-colored lithology \citep{RN23, RN4, RN70, RN914}. This corresponds to a shock stage C-S5 on the scale by \cite{RN146, RN138}. As the amount of post-shock heating increases, the silicates and metal start to melt at larger proportions and form whole rock melt pockets, corresponding to a C-S6 shock stage. Sometimes, the impact melt can retain a portion of finely brecciated unmolten silicates, as described above in the Chelyabinsk impact melt lithology case \citep{RN4, RN70, RN914}. The iron-metal and troilite are finely dispersed within the molten silicate material forming spherical blebs \citep{RN23, RN24}, and eutectic mixtures of metal and troilite are often present \citep{RN437, RN70, RN914}. At pressures over $\sim$70 GPa, a complete material melting can occur, which corresponds to a C-S7 shock stage \citep{RN146, RN138}.

\section{Methods}

\subsection{Spherical shock experiment}

The details of the spherical shock experiment are given in \cite{RN462}. Here we provide a brief summary. The experiments were carried out at the Russian Federal Nuclear Center (RFNC) -- Zababakhin All-Russia Research Institute of Technical Physics (former VNIIP) in Snezhinsk, Russian Federation. A 4-cm diameter sphere was prepared from a single piece of the Chelyabinsk light-colored lithology, encapsulated in a vacuum inside a 6-mm thick steel jacket made of KH18N9 steel, and then loaded with a spherically converging shock wave produced by explosives placed on the outside of the steel jacket. After the experiment, the sample was left to slowly cool back to an ambient temperature. Analogous experiments were previously carried out with the material of the Saratov L4 ordinary chondrite \citep{RN389} and the Tsarev L5 ordinary chondrite \citep{RN908}, as well as with the iron meteorites, Sikhote-Alin and Chinga \citep{RN909, RN910, RN912}.

The above described setup allows for a gradual increase in the peak shock pressures from the sample rim toward its center. While the detailed pressure and temperature ($p$-$T$) profiles were intentionally withdrawn by RFNC for security reasons, rough estimates are provided in \cite{RN446}. In order to get a better understanding of the $p$-$T$ conditions within the sample, we carried out additional shock modeling of the experiment using the iSALE code (see Sec.~\ref{sec:shock}).

\subsection{Mineral and geochemical investigations}

The optical microscopy in transmitted polarized light and reflected light was done on a thin section at the Institute of Geology and Geochemistry Ural Branch of the Russian Academy of Sciences (RAS) in Ekaterinburg, Russian Federation using an Olympus BX51 microscope and at the Faculty of Science of the University of Helsinki (UH) in Helsinki, Finland, using a Leica DM2500 P microscope with a DFC450 C camera. The scanning electron microscopy (SEM) in back-scattered electrons (BSE) was done at the Ural Federal University, Ekaterinburg, Russian Federation (UrFU) and RAS on carbon-coated, thick and thin sections. Compositional mapping and imaging in BSE samples was done at UrFU using a Carl Zeiss $\Sigma$igma VP FE (field emission) SEM and an INCA X-Max EDS (energy dispersive spectra) detector. The mineral phase identification and precise silicate composition measurements was done at RAS using a Cameca SX100 EMPA (Electron MicroProbe Analysis) at the 15keV and 20 nA current and a Bruker XFlash6 EDS detector calibrated with mineral standards.

The X-ray diffraction (XRD) was done at UH using a PANanalytical X'Pert3 Powder instrument from small powder quantities drilled out by a microdrill. The XRD pattern processing was done using HighScore Plus software. The Raman spectroscopy was done on a thin section at RAS using a Horiba LabRam HR800 Evolution employing a 633 nm excitation laser and 600 gr/mm grating. The X-ray microtomography (XMT) scans were collected at the Department of Physics, University of Helsinki with custom-built Nanotom 180 NF tomography Phoenix|X-ray Systems and Services (part of GE Measurement Systems and Solutions, Germany) equipment \citep{RN927}.

\subsection{Reflectance spectra measurements}

The reflectance spectra were measured in the ultraviolet, visible, near-infrared, mid-infrared, and far-infrared (UV, VIS, NIR, MIR, and FIR) range at UH and at the Planetary Spectroscopy Laboratory of the German Aerospace Center (DLR) in Berlin, Germany. Measurements were primarily performed on a rough-cut surface. Additional data were obtained from small powder quantities drilled out by a microdrill.

The measurements at UH were carried out over the range of 0.25--3.2 \textmu m using an OL 750 automated spectroradiometric measurement system by Gooch \& Housego, which is equipped with a polytetrafluoroethylene (PTFE) and gold integrating spheres, and with specular reflection traps under atmospheric conditions. The illumination was provided using a collimated deuterium (UV) or tungsten (VIS-NIR) lamps. Sample spectra were measured relative to the PTFE (UV-VIS) or the gold (NIR) standards.

The measurements at DLR were carried out over the range of 0.35--100 \textmu m using a Bruker Vertex80V FTIR (Fourier-transform infrared) instrument in bidirectional geometry in g. The illumination was provided using a collimated deuterium (UV), tungsten (VIS-NIR), or silicon carbide rod Globar (MIR-FIR) lamps. Sample spectra were measured relative to the Labsphere Spectralon\textregistered{} (UV-VIS) or the gold (NIR-MIR-FIR) standards.

\subsection{Shock modeling}
\label{sec:shock}

The Dellen version \citep{RN450} of the iSALE-2D (impact-SALE) shock physics code \citep{RN391} was used to simulate the spherical shock-recovery experiment. The code is based on the Simplified Arbitrary Lagrangian Eulerian (SALE) hydrocode solution algorithm \citep{RN431}. To simulate the hypervelocity impact processes in solid materials, SALE was modified to include an elasto-plastic constitutive model, fragmentation models, various equations of state (EoS), and multiple materials \citep{RN396, RN399}. More recent improvements include a modified strength model \citep{RN400}, a porosity compaction model \citep{RN391, RN401}, and a dilatancy model RN402. By applying cylindrical symmetry in the y-axis to an Eulerian numerical mesh, the set-up consisted of a half sphere of dunite material of strength properties from \cite{RN458} in order to simulate the Chelyabinsk meteorite material embedded in steel of strength properties from \cite{RN390}. The analytical equations of states (ANEOS) provided with iSALE\ were used for deunite and iron.

\begin{figure*}
\centering
  \includegraphics[width=17cm]{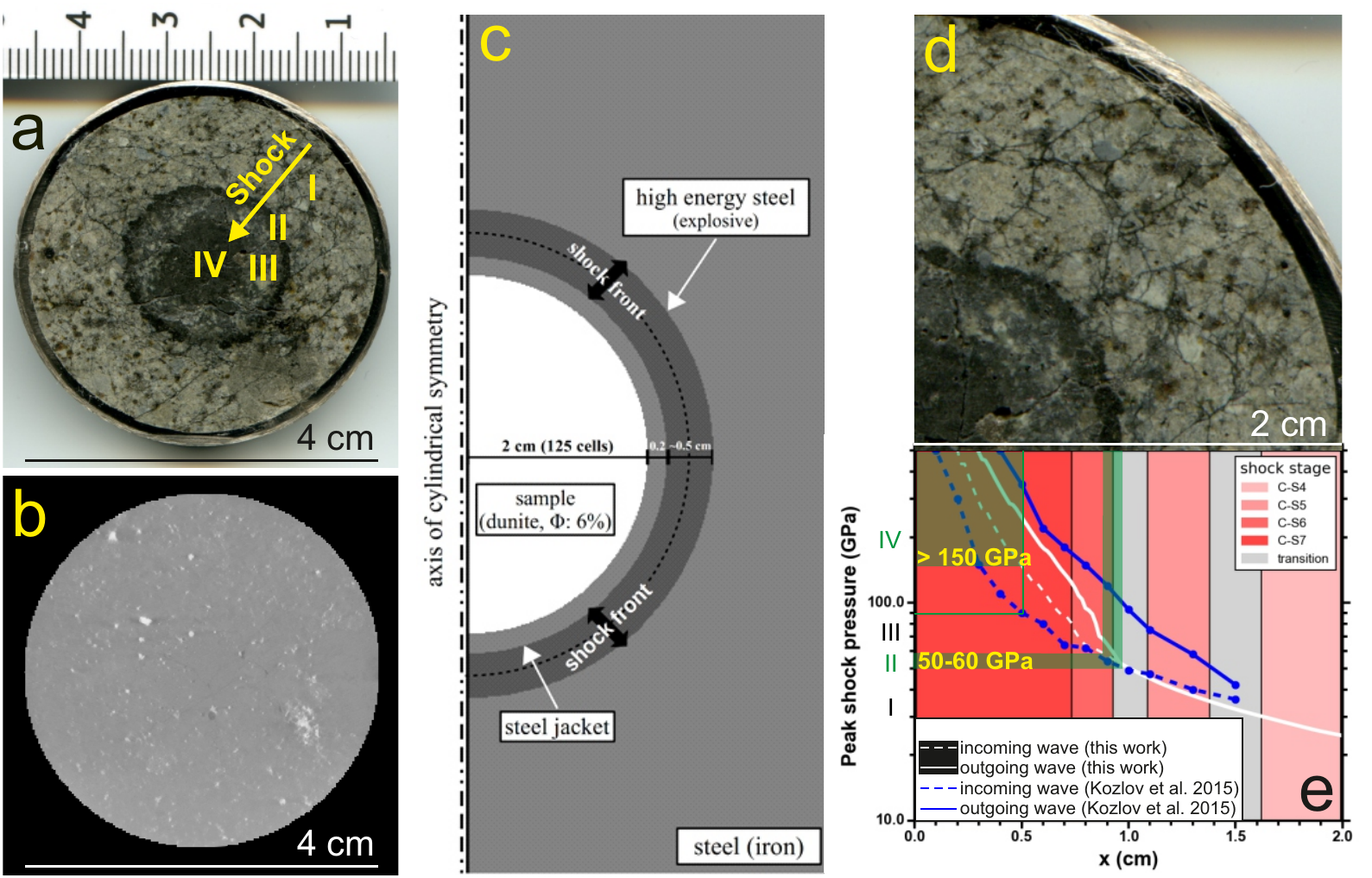}
  \caption{a: Spherically shocked Chelyabinsk meteorite sample. The arrow outlines the direction of increasing shock. The roman numbers distinguish the four different shock zones. b: XMT snapshot showing apparent lack or large metal and troilite grains (seen in bright colors) in the central zone IV. c: Set-up of the iSALE numerical model of the spherical shock-recovery experiment ($\Phi$ – porosity). d, e: Comparison of the experiment result (d, zoomed upper-right quadrant of a) to the iSALE modeling results of the spherical shock experiment (e) with the shock stage intervals from \cite{RN146, RN138}. The green areas in (e) outline the location of the shock-darkened zones II and IV, together with their pressure boundaries calculated from the inward shock wave peak shock pressures from our iSALE model. The blue-dashed and full lines correspond to inward and outward shock wave pressures from \cite{RN446}. The corresponding lower zone IV boundary from the \cite{RN446} model is $\sim$90 GPa as indicated by a green line.}
  \label{Fig1}
\end{figure*}

To simulate the detonation of the explosive surrounding the steel jacket and to generate the spherical shock wave, a portion of the iron material is loaded with 4 MJ/kg of specific internal energy (Fig.~\ref{Fig1}c), producing 25 GPa of pressure entering the sample with a pulse duration of $\sim$5 µs. The shock wave travels inward in the sample and outward in the iron case. Simulations with a void rather than an iron case surrounding the loaded steel did not affect the results. The model resolution is 126 cells per 2 cm of the meteorite sample radius (cell size of $1.613\!\times\!10^{-4}$ m). As the exact experiment parameters (initial pressure, shock pulse duration) were not revealed to us, we determined these by "reverse-engineering" the experiment setup by testing several initial pressure amplitudes (10--30 GPa) and shock pulse durations (3--5 \textmu s) in order to determine the above-mentioned initial parameters producing the best fit to the observed shock features in the sample.

\section{Results}

\subsection{Mineralogical and chemical investigations}

Following the shock experiment, the sphere with the meteorite sample was cut open in half. Four distinct roughly concentric zones can be macroscopically identified (Fig.~\ref{Fig1}a, modified from \cite{RN462}). In this publication, we number the zones in the direction of increasing shock from the outside toward the center as zones I--IV. Their boundaries are approximately located at 1 cm from the center, or at a radius of 0.5 R (zone I/II boundary), 0.9 cm or 0.45 R (II/III), and 0.5 cm or 0.25 R (III/IV). The original Chelyabinsk meteorite macroscopic texture is largely preserved in zone I. Zones II and IV are optically dark, while zone III is intermediate in its appearance. Numerous fractures and dark shock melt veins run through the sample.

\begin{figure*}[!htb]
\sidecaption
  \includegraphics[width=12cm]{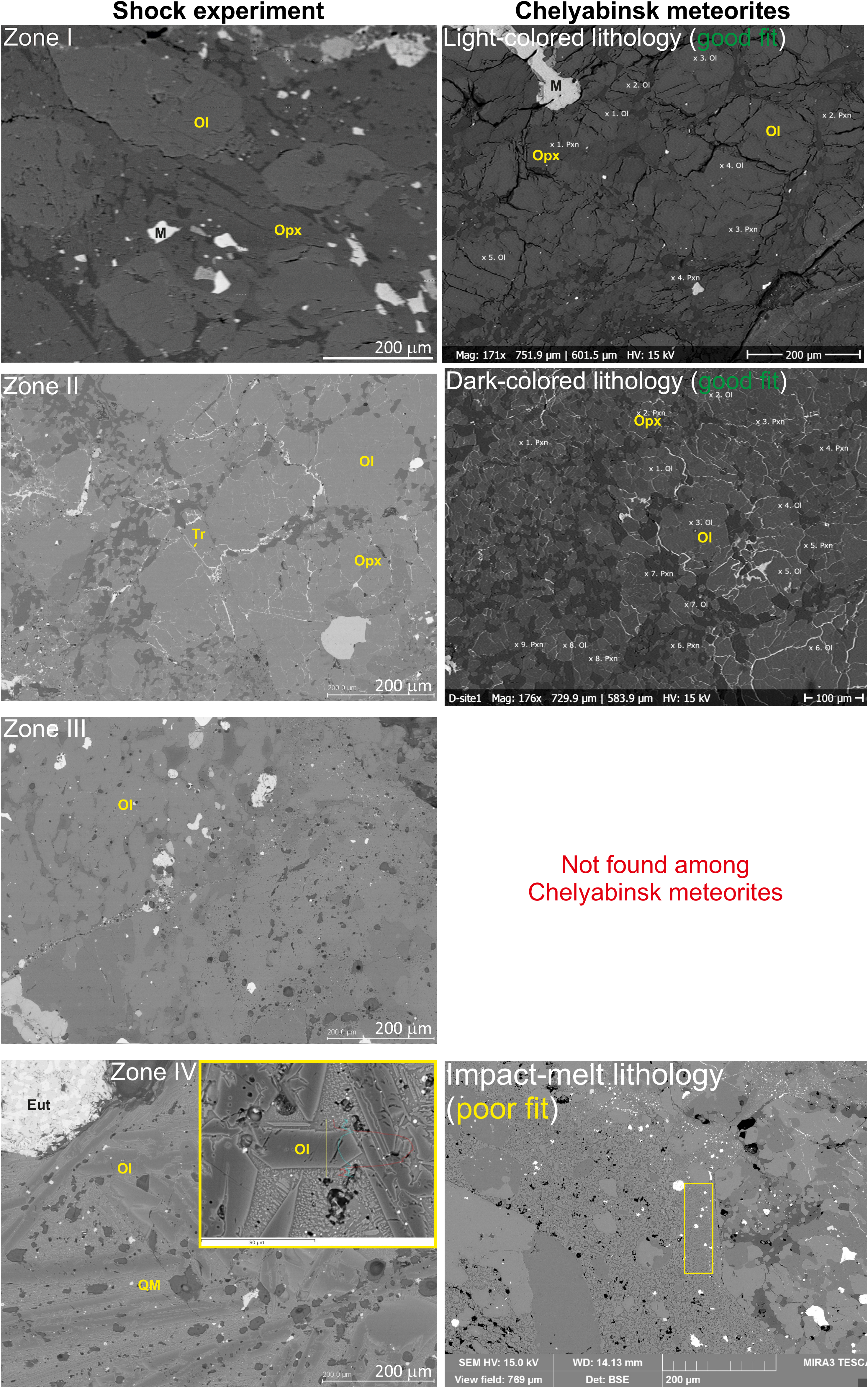}
  \caption{SEM-BSE images of the four zones together with their Chelyabinsk meteorite lithology equivalents. The inset in zone IV shows a detail of olivine skeletal structure and the metal and troilite redistribution into fine eutectic micro and nanoparticles (bright blebs), causing the observed optical darkening. The yellow box in the impact-melt lithology shows an area with needle olivine crystals in quenched melt, similar to those observed by \cite{RN914}. QM, Ol, Opx, and Tr  stand for quenched melt, olivine orthopyroxene, and troilite, respectively.}
  \label{Fig2}
\end{figure*}

Optical and electron microscopy reveal the following features (Figs.~\ref{Fig2} and \ref{FigS1}). Zone I is highly mechanically fractured. The olivine crystals show undulatory extinction and mosaicism. Numerous opaque shock veins and melt pockets are present. Their amount, together with the olivine mosaicism and occurrence of planar deformation features (PDFs), is higher compared to the original light-colored lithology and increases from the outer boundary toward zone II. Plagioclase is largely transformed into maskelynite. Thus, the Chelyabinsk material experienced slightly higher shock loading within zone I compared to its initial natural state (C-S3, locally C-S4) with an apparent shock gradient toward zone II. We classify its shock stage as C-S4, reaching C-S5 close to the boundary with zone II.

Zone II is opaque and thus not transparent in the optical microscopy. Reflected light and electron microscopy reveal the presence of a very fine network of troilite melt veins, protruding into fractures within silicate grains. Thus, zone II resembles the dark-colored lithology and we estimate that its shock stage is transitional between C-S5 and C-S6.

Zone III is characterized by highly brecciated olivine grains surrounded (or "floating") within silicate melt. Compared to zone II, the olivine grains are free of troilite melt. Plagioclase is shock-melted. Eutectic iron-troilite grains, abundant opaque shock veins, and melt pockets are present. Based on these characteristics, we assign this zone a shock stage of C-S6.

Zone IV is characterized by a recrystallized olivine phase in the form of needle-shaped crystals embedded in a quenched residual silicate melt. The olivine crystals are free of any observable shock effects and thus they crystallize after the shock loading from melt or vapor. Numerous voids as well as the iron-troilite eutectic droplets are present in the quenched melt. Due to the complete recrystallization, we assign this zone a shock stage of C-S7.

\begin{figure*}[!htb]
\centering
  \includegraphics[width=17cm]{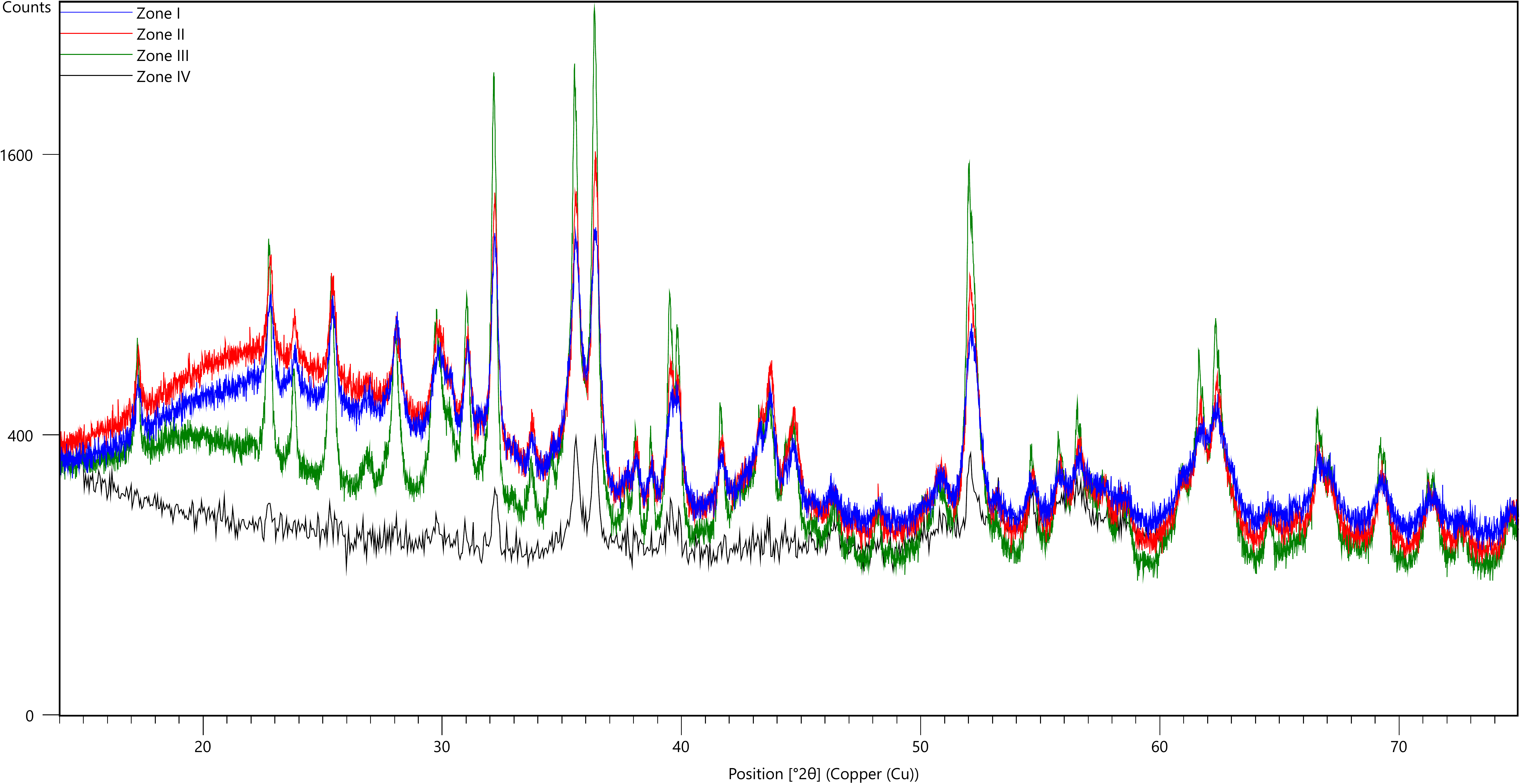}
  \caption{XRD patterns of zone I--IV powdered samples.}
  \label{Fig3}
\end{figure*}

The XRD results (Fig.~\ref{Fig3}) are noisy due to the small amount of a powdered sample used (especially in the sample from zone IV) and only qualitative comparison is possible. There is no apparent difference in the diffraction pattern among samples from zones I--IV, thus indicating no major mineral changes due to the shock. The observed peaks can be assigned to the following three most abundant constituents of the ordinary chondrites: olivine, orthopyroxene, and plagioclase (Fig.~\ref{FigS2}).

The EMPA analysis was done on olivine and orthopyroxene from all four zones in order to verify possible changes in their chemistry (e.g. Fe/Mg ratio). The results are outlined in Table 1. The composition of olivine and orthopyroxene in zones I-III is consistent (Fa $\sim$30, Fs $\sim$25), while the needle olivine in zone IV has a distinct iron-poor (Fa $\sim$17-18) composition with the surrounding quenched melt, which is Fe-rich. Moreover, the olivine needle in this zone shows a distinct zonation with Fe-enriched rims and a Mg-enriched center (Fig.~\ref{FigS3}).

Raman spectra were measured from olivine grains from all zones. The spectra exhibit intense Raman bands B1 ($\sim$820 cm$^{-1}$) and B2 ($\sim$850 cm$^{-1}$), which are typical for olivine (Table~1, Fig.~\ref{FigS4}). The positions of the bands are influenced by the chemical composition \citep{RN920} and cannot be used as a reliable shock indicator. The band width, however, can be interpreted as a rough proxy to the shock pressure \citep{RN915}. In Table~1, we present  the full width at half maximum (FWHM) for both bands as well as the Fa content calculation based on the band position, following the method in \cite{RN921}. The Raman-derived olivine composition results are close to the EMPA results, and they indicate a rather uniform composition of zones I--III and an Fa depletion in zone IV. Even so, it is important to note that a Raman-based calculation is less precise compared to an EMPA analysis, especially the calculation based on B1. The FWHM results of both bands indicate a trend in band broadening due to a possible increase in the relative shock effect in a sequence of IV-III-I-II.

The XMT imaging reveals an apparent lack of large metal and troilite particles in zone IV (Fig.~\ref{Fig1}b). This is a result of the metal and troilite redistribution into eutectic micro and nanoparticles as revealed by the high-resolution SEM imaging (Figs.~\ref{Fig2}, and \ref{FigS3}). A similar phenomenon was reported earlier by \cite{RN24}, which is also consistent with almost identical grain density and magnetic susceptibility values among all three analogous natural Chelyabinsk light-colored, dark-colored, and impact melt lithologies \citep{RN4}, thus implying similar overall
mineralogy and metal content.

\subsection{Reflectance spectra}

\begin{figure*}[!htb]
\centering
  \includegraphics[width=17cm]{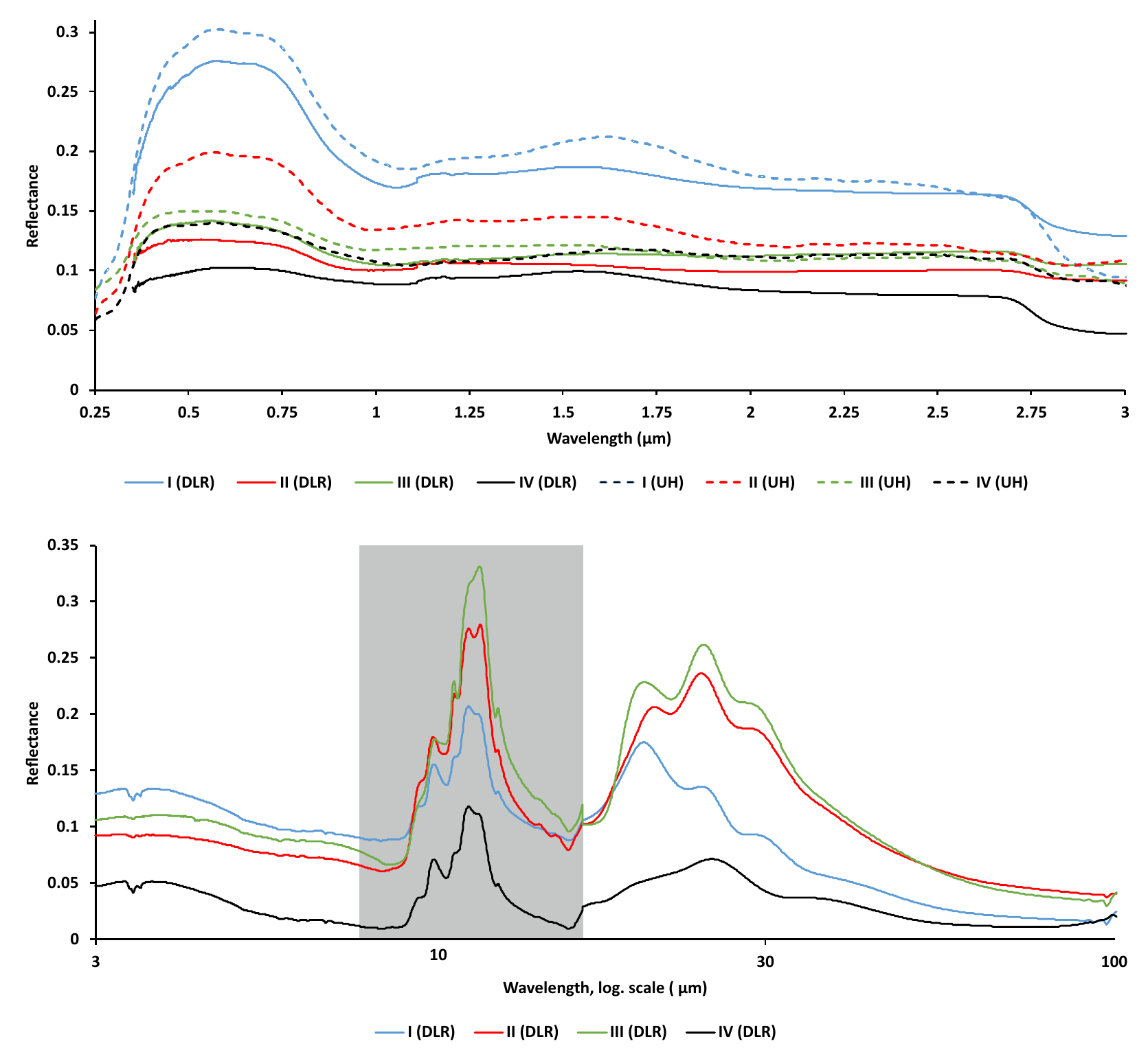}
  \caption{Comparison of the UV-VIS-NIR bidirectional (DLR) and hemispherical (UH) reflectance (top) and the MIR bidirectional reflectance measured at DLR (bottom) of the four zones from the rough saw-cut sample surface. The area in gray is shown in detail in Fig.~\ref{Fig5}.}
  \label{Fig4}
\end{figure*}

Fig.~\ref{Fig4} (top) shows a comparison of the bidirectional (DLR) and the hemispherical (UH) reflectance in the UV-VIS-NIR (0.25-3 \textmu m) region as measured from a rough saw-cut sample surface. The presence of 1-\textmu m olivine \citep{RN136} and 1- and 2-\textmu m orthopyroxene \citep{RN137} absorption bands can be seen in all four zones. The order of the spectral curves in the UV-VIS-NIR region follows the visual brightness (Fig.~\ref{Fig1}a) in which zone I is the brightest, followed by zones III and II, and zone IV is the darkest one. The hemispherical reflectance is slightly higher than the bidirectional reflectance, and the absorption bands are more pronounced. The drop in the reflectance at 2.7 \textmu m is caused by absorbed terrestrial water.

\begin{figure*}[!htb]
\centering
  \includegraphics[width=17cm]{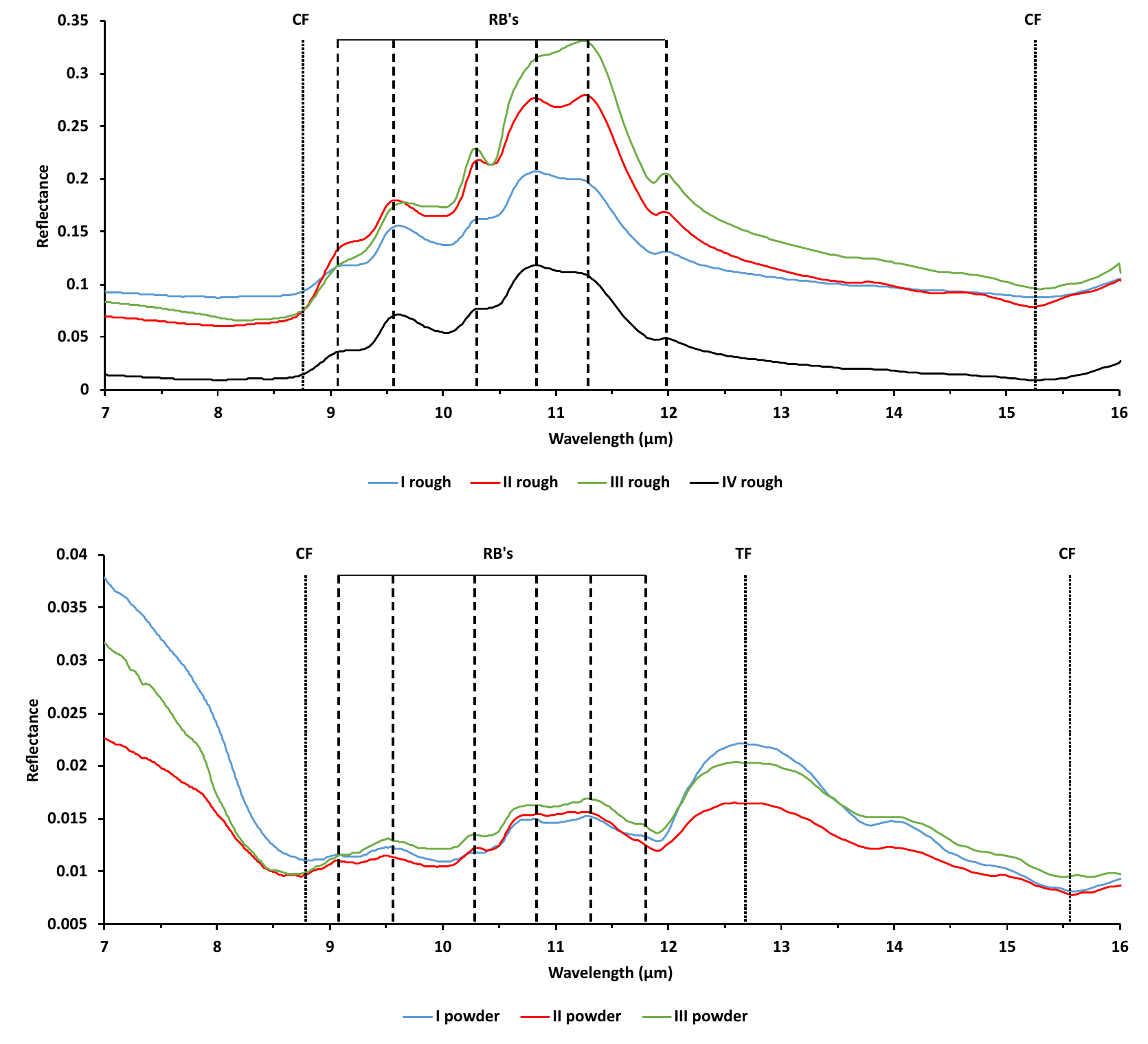}
  \caption{Comparison of the MIR bidirectional (DLR) reflectance of the four zones measured on a saw-cut surface (top) and powder (bottom). CF, RB, and TF stand for the Christiansen feature, reststrahlen bands, and the transparency feature, respectively.}
  \label{Fig5}
\end{figure*}

Fig.~\ref{Fig4} (bottom) shows the MIR-FIR bidirectional (DLR) reflectance as measured on the rough saw-cut surface. The MIR reflectance follows the same albedo order as UV-VIS-NIR up to the primary Christiansen feature at 8.7 \textmu m. At higher wavelengths in the Si-O reststrahlen bands region, the reflectance order changes with zones II and III, which are brighter than zones I and IV. The gray zone in Fig.~\ref{Fig4} (bottom) highlights the 7--16 \textmu m region, which is the zoomed in on in Fig.~\ref{Fig5} (rough saw-cut surface and powdered sample). The powdered sample measurement of the zone IV sample is missing due to the extremely small amount of available powdered material. Fig.~\ref{FigS5} displays normalized data from Fig.~\ref{Fig4}.

The comparison of the powdered sample spectra to the one obtained from the rough saw-cut surface reveals the following trends. The overall reflectance of the powdered sample is an order of magnitude lower compared to the rough surface one. The reststrahlen bands in both samples show similar positions at approximately 9.1, 9.5--9.6, 10.3, 10.8, 11.3, and 11.8--12 \textmu m, they are dominated by olivine \citep{RN424, RN421, RN429}, and there is the possible presence of orthopyroxene \citep{RN422}. The amplitudes of the reststrahlen bands are higher in the rough surface sample. The transparency feature at 12.7 \textmu m is only observed in the powdered sample \citep{RN430}. The primary Christiansen feature at 8.7 \textmu m is more pronounced in the powdered sample \citep{RN424}, while the secondary one at 15.6 \textmu m \citep{RN429} is of a low amplitude in both samples. Three additional major broad Si-O reststrahlen bands are observed in the FIR region at 20, 24, and 29 \textmu m, which belong to olivine \citep{RN421, RN429}.

\subsection{Shock modeling}

Fig.~\ref{Fig1}e shows the calculated peak shock pressures from the inward wave culminating at 3.3 \textmu s and the subsequent decaying outward wave at a 5.6 \textmu s model run time. The initial pressure at the rim of the spherical sample is $\sim$25 GPa and rises to pressures above 400 GPa at the center due to the converging geometry of the primary shock wave. When the pressure culminates in the center of the sample, an outward diverging secondary shock wave causes a secondary rise in the shock pressures due to the superposition with the inward propagating primary shock wave, similarly to what was observed by. The secondary diverging shock wave continuously decays  with no reflections back at the steel interface with the sample.

The shock stage intervals are determined by comparing the inward wave peak shock pressures (as an inward wave delivers approximately 90\% of the total shock energy) to the pressure ranges given in \cite{RN146, RN138}, which  are outlined in Fig.~\ref{Fig1}e with the following distribution expressed in the sample radius: C-S4 extends to $\sim$0.8--1 R, C-S5 to $\sim$0.55--0.7 R, C-S6 to $\sim$0.35--0.45 R, and C-S7 to $\sim$0--0.35 R. The location of the shock stage regions, as derived from the iSALE model, agrees with the observations of the shock features in the sample with the exception of a slight offset in the C-S6 and C-S7 boundary location toward lower pressures. The possible reasons for this are outlined in the following section. The shock-darkened zone III corresponds to the pressure interval of 50--60 GPa in the iSALE model.

\section{Discussion}

There are both strong similarities as well as significant differences observed among the shock effects found in the naturally occurring Chelyabinsk meteorites and our experimentally produced samples. Zone I maintains a similar overall appearance as the original light-colored lithology, despite the additional shock loading caused by the shock experiment. The shock effects in silicates seem to be slightly stronger (e.g. pronounced mosaicism, PDFs, and maskelynite), and the abundance of the opaque shock veins and melt pockets is higher resulting in an overall increase in the shock stage classification from C-S4 in the outer part to C-S5 in the inner part of zone I compared to C-S3 of the original Chelyabinsk light-colored lithology.

Zone II resembles the Chelyabinsk dark-colored lithology. Similar to the naturally occurring dark-colored Chelyabinsk meteorites, the darkening is mainly attributed to the partial melting of troilite and the subsequent injection of troilite melt into the cracks within silicate grains. This fabric is typical for the transition between shock stages C-S5 and C-S6 \citep{RN146, RN458, RN138}. Compared to the naturally occurring dark-colored lithology, the extent of troilite melting seems to be slightly lower in our experiment since some of the silicate grains are still partly transparent in the optical microscopy (Fig.~\ref{FigS1}). This can be explained due to a shorter timescale of the shock-induced heating (shorter duration, faster cooling rates) in the experiment compared to the asteroid collisions. The minimal thickness of zone II, which is approximately 1 mm (Fig.~\ref{Fig1}d), is a surprising finding, corresponding to a narrow 10 GPa interval in the peak-shock pressures as derived from the iSALE shock model (Fig.~\ref{Fig1}e).

The presence of the subsequent zone III with its optically brighter appearance and virtually missing shock darkening effects is a surprise. Our initial expectation based on the review of shock effects in chondrites was that since the onset of the shock darkening related to troilite melt, the material retains its dark appearance and will transition directly into whole-rock melt with finely dispersed eutectic blebs of metal and troilite, as observed in the Chelyabinsk meteorites. An explanation for the sudden disappearance of shock darkening can be found by taking a closer look at the microstructure. An onset of significant silicate melting is observed. The remnant silicate grains are highly fractured and surrounded by, or floating in, a silicate melt. The troilite and the metal form eutectic grains within the silicate melt, but they do not enter the remnant silicate grains. We attribute the sudden lack of shock darkening to mainly the above-described presence of silicate melt along grain boundaries. Metal and troilite as well as silicate melts are known to be immiscible \citep{RN436}. Thus, the silicate melt creates an isolating layer surrounding the remnant silicate grains and protects their interiors from an injection of the molten metal or troilite. To our knowledge, no equivalent for zone III has been reported among the recovered Chelyabinsk meteorites. This phenomenon also narrows the pressure range of the troilite-related shock darkening to only approximately 10 GPa ($\sim$50--60 GPa in Chelyabinsk case) and it places constraints on the volume of naturally produced shock-darkened material during asteroid collisions.

The material within zone IV seems to have reached a complete melting or vaporization (shock stage C-S7). New needle-shaped 0.5--1 mm-long crystals of olivine, which are free of any shock effects, are observed together with residual quenched silicate melt, vesicles, as well as eutectic metal and troilite grains. The needle-shaped olivine crystals are approximately two to ten times larger compared to the needle crystals observed in the natural impact melt lithology (Fig.~\ref{Fig2} or \cite{RN914}). The fine texture and redistribution of the eutectic metal and troilite grains within the abundant melt contribute to the dark appearance of this innermost material. At first glance, this zone roughly resembles the Chelyabinsk impact-melt lithology. The latter one, however, sometimes contains clasts of the original unmolten material, which is not observed in zone IV (Fig.~\ref{Fig2}). Zone IV is closer in its nature, for example, to the shock-melted material found in Krymka LL3.1 chondrite \citep{RN933}.

Olivine crystals in zone IV are highly skeletal (Figs.~\ref{Fig2} and \ref{FigS3}) with common needle or prismatic forms, which were classified as “hopper olivine” to “chain olivine” \citep{RN891} or as “hopper” to “baby swallowtail” from \cite{RN885} terminology. In the dynamic crystallization experiments \citep{RN885, RN862}, hopper and baby swallowtail olivines crystallize in a wide range of cooling rates, 188–1552°C/h, the degree of undercooling being 50--80°C. The relatively fast cooling rates are also supported by the unequilibration (Fe-Mg zonation) observed in the olivine crystals.

The surprising finding is the indication of an out-of-order (IV-III-I-II) shock trend revealed by the olivine Raman band broadening. Zone IV represents post-shock recrystallized material. Thus, it is free of any shock features and has the narrowest olivine Raman bands. The remnant zone III olivines were shocked to levels of CS-6 or $\sim$50--60 GPa. However, they may have partly healed or recrystallized through the significant material heating and partial melting, and they therefore exhibit relatively low Raman band broadening compared to zones I and II, despite the high shock load. The olivines in zone II show the strongest Raman band shock broadening as they experienced the highest shock load without the above-described healing effects.

The UV-VIS-NIR reflectance spectrum of zone I is consistent with the previous measurements of the light-colored Chelyabinsk lithology \citep{RN4, RN61} with moderate reflectance values and the presence of 1 and 2-\textmu m silicate absorption bands. Zones II and IV are significantly darker and are spectrally similar to each other as well as to the Chelyabinsk dark-colored and impact-melt lithologies \citep{RN4, RN61}. The darkening is caused by metal and troilite redistribution (Figs.~\ref{Fig2} and \ref{FigS1}) from large distinct grains into fine melt veins (zone II) or fine eutectic particles (zone IV), which is in line with the previous shock-darkened meteorite studies as reviewed in the Introduction. The silicate absorption bands are, however, not completely erased as is often found to be the case in the Chelyabinsk dark-colored and impact-melt lithologies. This can be explained by the lower extent of troilite melt migration to silicates due to the shorter time span of our experiment (zone II) or the presence of newly crystalized olivine (zone IV). The brightness and the UV-VIS-NIR spectrum of zone III is intermediate compared to zone I as well as zones II and IV.

The MIR-FIR spectrum is similar to the one observed in natural Chelyabinsk meteorites by \cite{RN62}. Generally, the reststrahlen band positions are consistent among all four zones as well as between the rough surface and the powdered material. We do not observe the loss of the 10.8/11.3 \textmu m twin-peak or other features as a result of shock or melting as observed by \cite{RN62}. This can be due to the fact that our experimentally shocked sample does not show the extensive presence of glassy structures, which are found in the Chelyabinsk impact melt lithology. Zones I-III still contain a significant fraction of crystalline material, while zone IV consists of well crystallized new olivine. Even so we reached complete melting or even vaporization in our experiment, the cooling in the sample center was slow enough for new olivine to crystallize. The amplitude of the reststrahlen bands as well as the overall reflectance are higher in the raw cut surface, while the transparency and the Christiansen features are more pronounced in the fine powdered material, which is in agreement with \cite{RN430, RN424, RN62}. No correlation is observed between the amplitude of the individual reststrahlen bands and the level of the shock. The important finding is that the MIR-FIR features in zones II and IV are not erased or obstructed by the shock darkening effects, as is the case of the VIS-NIR features. Thus, the MIR-FIR spectroscopy can be used to detect compositions of shocked chondritic planetary materials.

The mineralogical (XRD, Raman spectroscopy) and the chemical (SEM-EDS) analysis methods reveal no major mineral changes among zones I--III due to shock loading, and the derived composition is similar to the one observed on the natural Chelyabinsk meteorite samples. The silicate chemistry of zone IV, however, deviates from Chelyabinsk meteorites. The newly crystalized olivine (Fa 17--18) in zone IV is iron-poor, which is typical of H chondrites rather than LL. In contrast, the surrounding residual quenched silicate melt is iron rich, suggesting Fe fractionation during the cooling and the crystallization of zone IV silicates.

Although the results of the numerical model help us to understand the overall shock distribution in our experiment, uncertainties exist, especially in the highly shocked zone close to the interior. They are caused by the following reasons: (1) The exact conditions of the shock experiment, such as the initial pressure at the steel-meteorite interface and the pulse duration, were not revealed to us, and (2) due to the use of the cylindrical symmetry, the grid approaches singularity at the center and suffers from a relative loss of resolution. Thus, the results at distances approximately < 0.25 R should be interpreted with caution.

The presented model with the peak shock pressure distributions is the closest proxy to the observed shock effects within a sample (e.g. extent of shock zones), and it is in general agreement with another model by Kozlov, Degtyarev et al. (2015) at a radius > 0.8 mm for the converging wave (Fig.~\ref{Fig2}e). The smaller separation of the converging and diverging shock waves in our model compared to \cite{RN446} at a radius < 0.8 mm can be explained by reverberation effects \citep{RN447, RN448, RN449}. The zone III and IV transition also occurs at higher pressures ($\sim$150 GPa) based on our model compared to results by \cite{RN446} ($\sim$90 GPa) due to the above-mentioned discrepancies. 

\section{Conclusions}

The spherical shock experiment with the Chelyabinsk meteorite allows us to observe how the ordinary chondrite material reacts to a gradually increasing shock load. A moderate shock (up to $\sim$50 GPa in the case of LL5 Chelyabinsk material) results in an increasing abundance of shock features in minerals and melting, localized mainly in the shock veins and melt pockets (zone I). These processes do not have a significant impact on the reflectance spectra. An increase in the shock load results in the significant generation of troilite melt and its flow into cracks within olivine and pyroxene grains (zone II). The texture resembles the Chelyabinsk dark-colored lithology with a significant spectral darkening in the UV-VIS-NIR region and a suppression of the silicate 1 and 2-\textmu m absorption bands. This process, however, occurs in a surprisingly narrow pressure interval of $\sim$10 GPa. At higher pressures (over 60 GPa in the Chelyabinsk case), the silicates start to melt along the grain or the crack boundaries and the silicate melt shields the residual grains from an infusion of  troilite melt (zone III). This is due to the immiscibility of silicate and troilite melts. The material restores its bright appearance, to a large extent, with reflectance spectra similar to zone I. The next and final step in the shock metamorphism is the complete melting of material (at pressures over $\sim$150 GPa in the Chelyabinsk case). This is represented by zone IV in our experiment and is characterized by newly crystallized Fa-poor olivine needle crystals, which are embedded in a quenched Fe-rich residual melt. In our experiment, the melt zone does not contain any brecciated original material, which is present in naturally occurring Chelyabinsk impact-melt lithology and the proportion of the needle crystals versus melt is higher. The quenched melt contains numerous fine metal and troilite eutectic blebs, causing spectral darkening similar to zone II.

The important finding is the presence of the two distinct shock darkening pressure mechanisms in ordinary chondrite material with characteristic material fabric and distinct pressure regions. These two regions are separated by a pressure interval where no darkening occurs. Thus, the volume of the darkened material produced during asteroid collisions may be somewhat lower than assumed from a continuous darkening process. While the darkening mainly affects the UV-VIS-NIR region and 1 and 2-\textmu m silicate absorption bands, it does not significantly affect the silicate spectral features in the MIR region. These are more affected by material roughness. MIR observations especially have the potential to identify darkened ordinary chondrite material with an otherwise featureless UV-VIS-NIR spectrum.

\begin{acknowledgements}

This research
 is supported by Academy of Finland project no. 293975, NASA SSERVI Center for Asteroid and Lunar Surface Science (CLASS), the Project No. 0836-2020-0059 of the Ministry of Science and Higher Education of the Russian Federation, the Act 211 of the Government of the Russian Federation, agreement no. 02.A03.21.0006, RFBR project no. 18-38-00598, the theme of state assignment of IGG UB RAS number AAAA-A19-119071090011-6, and within institutional support RVO 67985831 of the Institute of Geology of the Czech Academy of Sciences. We would like to thank Andreas Morlok for his constructive review comments on the manuscript.

\end{acknowledgements}

\bibliographystyle{aa}
\bibliography{References}

\begin{appendix}

\section{Data archive}

All measured data as well as full resolution images can be found at \url{https://dx.doi.org/10.5281/zenodo.3584942}.

\begin{table}[!htb]
\caption{EMPA and Raman analysis of olivine and orthopyroxene of the four shocked zones. We note that ol, opx, s.d., and no. stand for olivine, orthopyroxene, the standard deviation, and the number of analyzed grains, respectively.}
\label{table1}
\centering
\begin{tabular}{l r r r r}          
\hline\hline
Zone & I & II & III & IV \\
\hline
 & & & & \\
\hline
EMPA (ol and opx) &   &    &     &    \\
\hline
Fa average & 30.2       & 29.7 & 30.0 & 18.6 \\
Fa median  & 30.1 & 29.5 & 30.0 & 17.2 \\
s.d.       &  0.6 &  0.9 &  0.4 &  2.1 \\
no.        &   20 &   17 &   26 &   18 \\
\hline
Fs average & 25.2 & 24.7 & 25.1 &  N/A \\
Fs median  & 25.2 & 24.5 & 24.9 &  N/A \\
s.d.       &    0.4 &  0.6 &  0.6 &  N/A \\
no.        &   16 &   10 &   11 &  N/A \\
\hline
 & & & & \\
\hline
Raman (ol) & & & & \\
\hline
B1 position average & 819.6 & 818.7 & 819.4 & 819.8 \\
s.d. & 0.5 & 0.3 & 0.9 & 0.4 \\
B1 FWHM average & 15.9 & 16.8 & 15.6 & 14.2 \\
s.d. & 0.8 & 1.7 & 1.8 & 0.9 \\
B2 position average & 850.1 & 849.1 & 850.2 & 851.2 \\
s.d. & 0.2 & 0.5 & 0.4 & 1.2 \\
B2 FWHM average & 20.2 & 21.0 & 18.6 & 17.2 \\
s.d. & 0.5 & 2.0 & 0.9 & 1.4 \\
no. & 5 & 3 & 3 & 3 \\
Fa average (from B1) & 33.8 & 41.3 & 35.4 & 31.9 \\
s.d. & 4.7 & 3.2 & 8.4 & 3.8 \\
Fa average (from B2) & 24.4 & 31.3 & 23.8 & 18.8 \\
s.d. & 1.0 & 2.8 & 2.0 & 5.8 \\
\hline
\end{tabular}
\end{table}

\section{Supplementary figures}

\begin{figure*}[hb]
\centering
  \includegraphics[width=17cm]{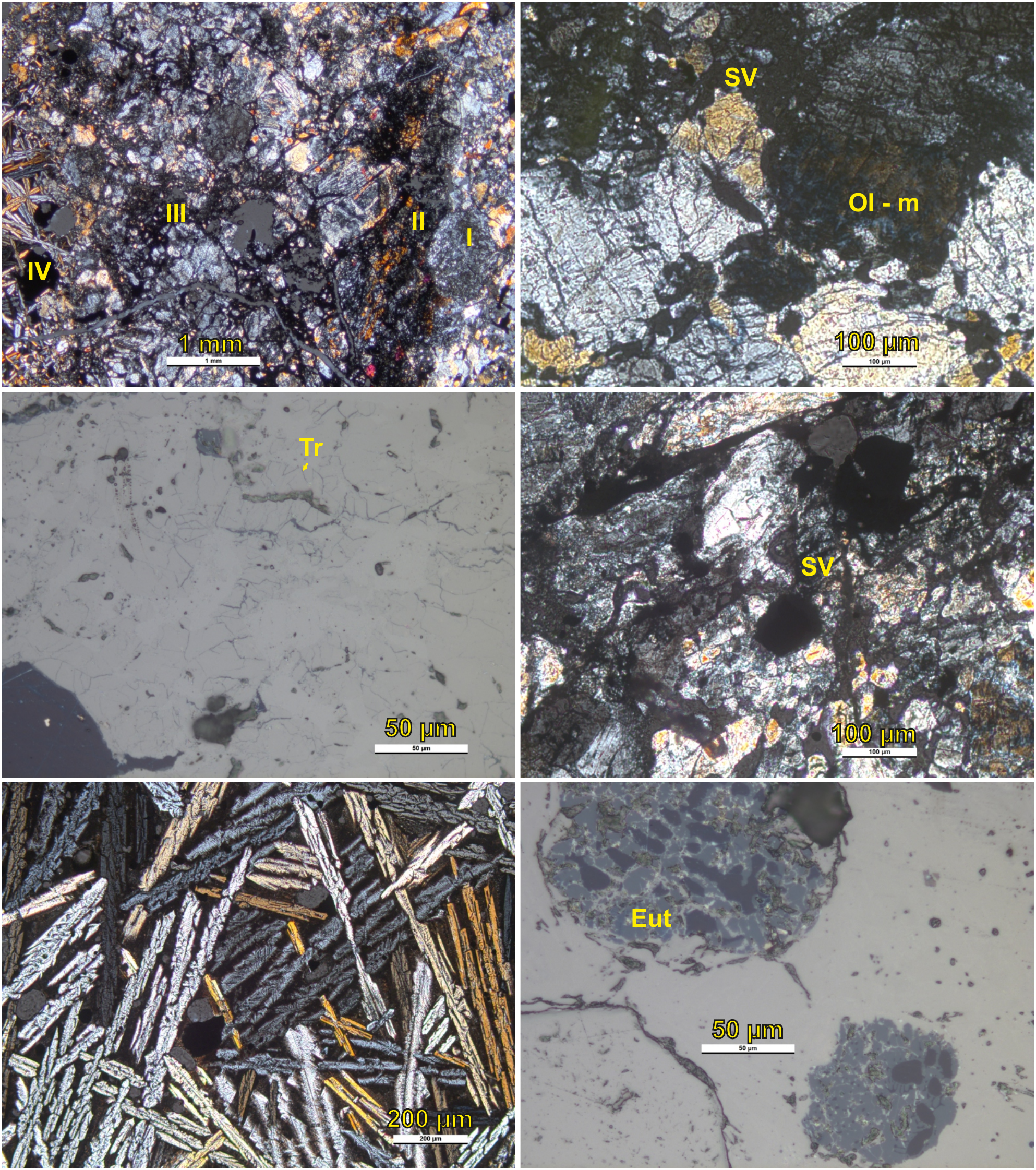}
  \caption{Microscope snapshots of the four shocked zones. General overview (top, left) in cross-polarized light with labeled zones. Zone I (top, right) in cross-polarized light. Ol - m is an olivine grain in extinction position showing mosaicism, SV indicates an opaque shock vein. Zone II (middle, left) is in reflected light. Tr with an arrow indicates darker troilite melt veins intruding silicates. Zone III (middle, right) is in cross-polarized light. Remnants of silicate grains are surrounded with silicate melt, SV indicates an opaque shock vein. Zone IV is in cross-polarized (bottom, left) and reflected (bottom, right) light. New needle-sized olivine crystals with sharp extinction and eutectic metal and troilite grains (Eut) are visible.}
  \label{FigS1}
\end{figure*}

\begin{figure*}[hb]
\centering
  \includegraphics[width=17cm]{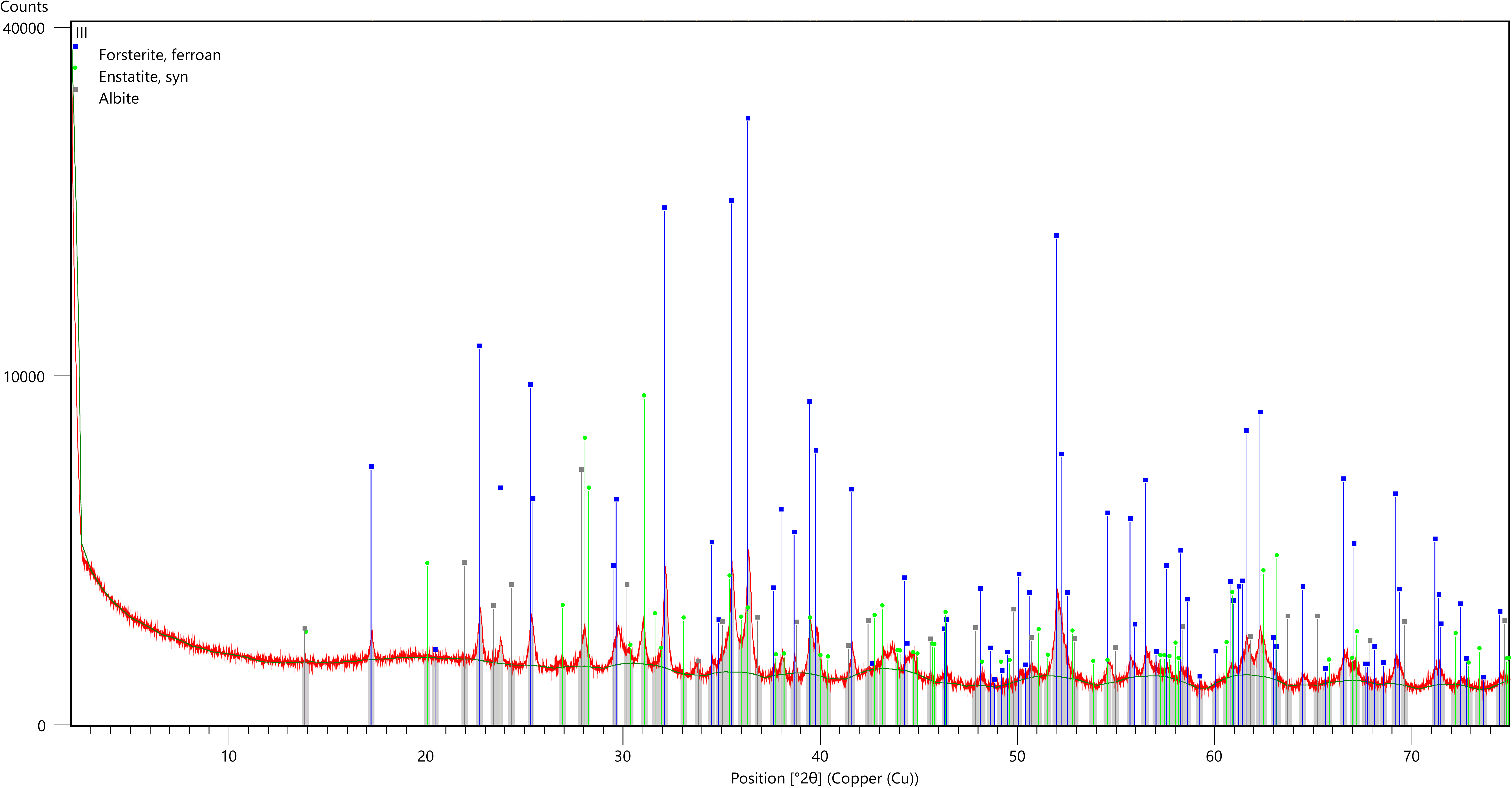}
  \caption{XRD pattern of zone III powdered sample (bottom) with diffraction patterns of olivine (forsterite), orthopyroxene (enstatite), and plagioclase (albite).}
  \label{FigS2}
\end{figure*}

\begin{figure*}[hb]
\centering
  \includegraphics[width=17cm]{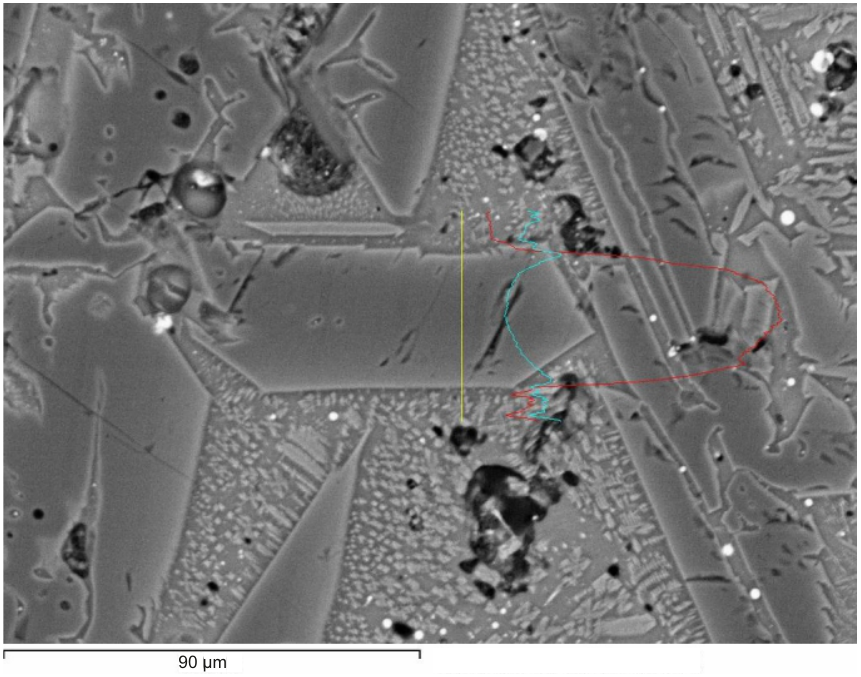}
  \caption{Detail of zone IV with skeletal olivine crystals and apparent metal and troilite redistribution into fine eutectic micro and nanoparticles (bright blebs) causing the observed optical darkening. The profile across the olivine crystal shows the distribution of Fe and Mg. Yellow marks the profile line. Blue, and red correspond to the amount given in detector counts for the Fe, and Mg, respectively.}
  \label{FigS3}
\end{figure*}

\begin{figure*}[hb]
\centering
  \includegraphics[width=17cm]{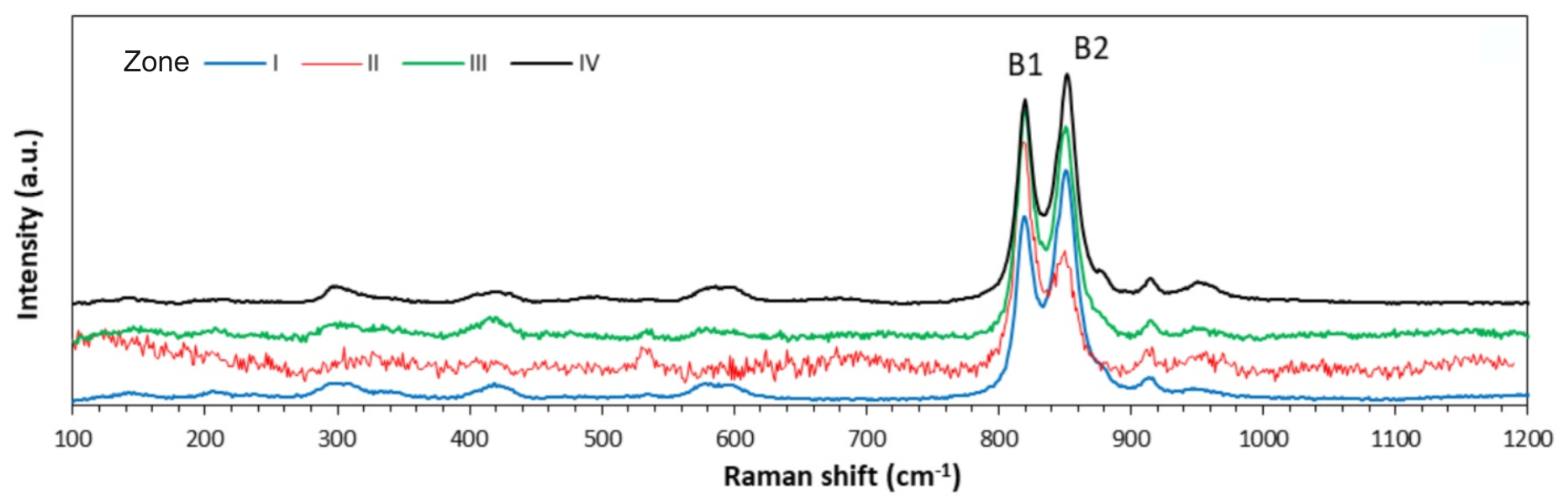}
  \caption{Example of Raman spectra of all four zones with positions of olivine B1 and B2 Raman bands.}
  \label{FigS4}
\end{figure*}

\begin{figure*}[hb]
\centering
  \includegraphics[width=17cm]{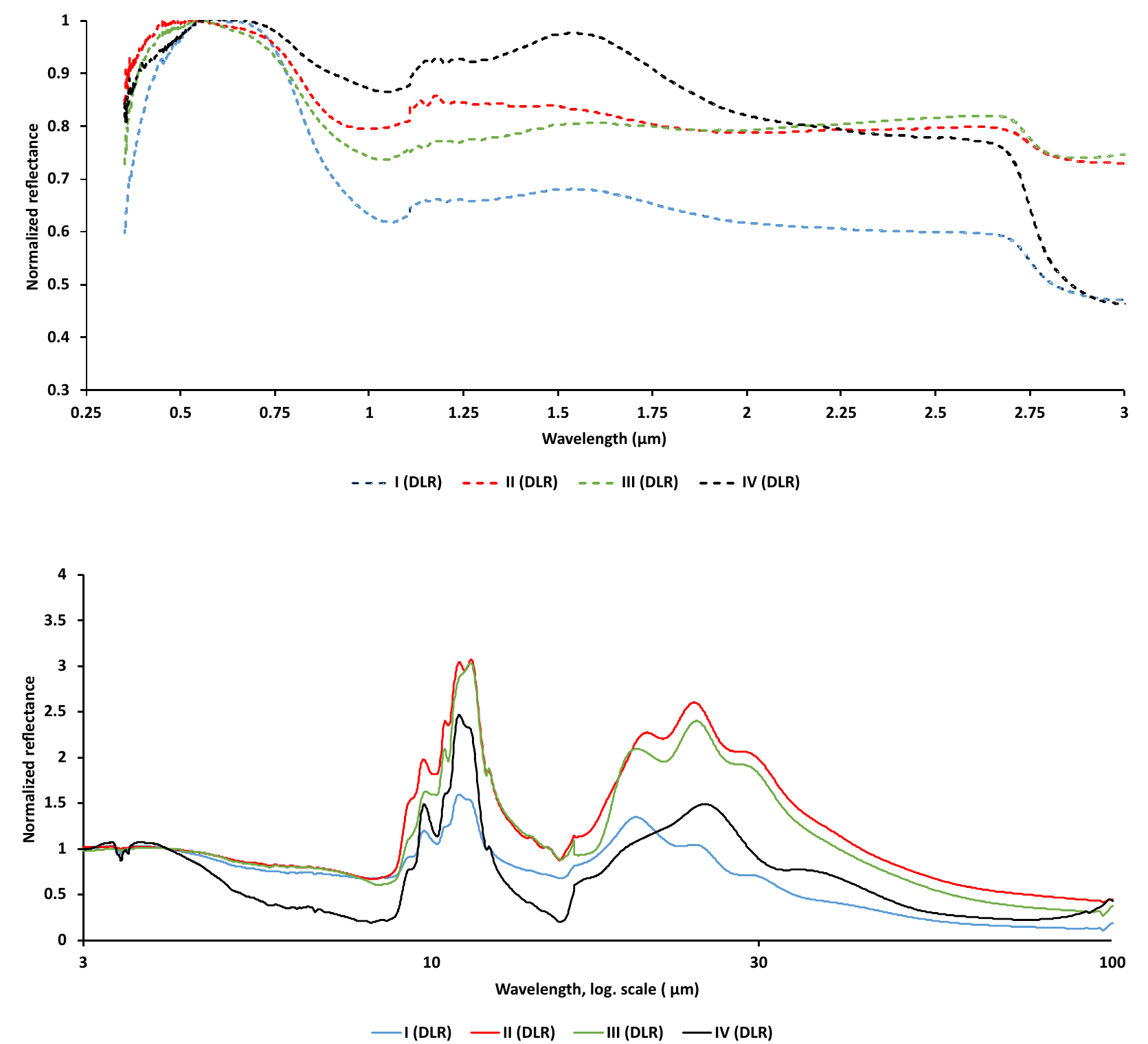}
  \caption{Normalized UV-VIS-NIR (top, normalized to unity at 0.55 \textmu m) and MIR (bottom, normalized to unity at 4 \textmu m) bidirectional reflectance measured at DLR of the four zones from the rough saw-cut sample surface.}
  \label{FigS5}
\end{figure*}

\end{appendix}

\end{document}